Running head: Reversible Watson-Crick Automata

Title: Reversible Watson-Crick Automata

Authors: Kingshuk Chatterjee[1], Kumar Sankar Ray(corresponding author)[2]

Affiliations: [1,2]Electronics and Communication Sciences Unit, Indian Statistical Institute, Kolkata-108

Address:    Electronics and Communication Sciences Unit, Indian Statistical Institute, Kolkata-108

Telephone Number: +918981074174

Fax Number: 033-25776680

Email:ksray@isical.ac.in

# Reversible Watson-Crick Automata

*Kingshuk Chatterjee[1], Kumar Sankar Ray[2]*

*Electronics and Communication Sciences Unit, Indian Statistical Institute, Kolkata-108.*

[1]kingshukchaterjee@gmail.com  [2]ksray@isical.ac.in

*Abstract:* Since 1970, reversible finite automata has generated interest among researchers, but till now we have not come across a model of reversible read only one-way finite automata which accept all regular languages, In this paper, we introduce a new model of one-way reversible finite automata inspired by the Watson-Crick complementarity relation and show that our model can accept all regular languages. We further show that our model accepts a language which is not accepted by any multi-head deterministic finite automaton.

*Keywords: non-deterministic Watson-Crick automata, deterministic Watson-Crick automata, reversible finite automata, reversible multi-head finite automata, reversible Watson-Crick automata.*

## 1. INTRODUCTION

The first study on reversible automata was done in the seventies [1], when Bennet introduced the concept of reversible Turing machines. He showed that the computational power of reversible Turing machines is same as that of Turing machines. This result encouraged researchers to explore the reversibility property in more restricted automata and in this endeavor Morita [2] explored the concept of two-way multi-head reversible automata and showed that it has the same computational power as two-way deterministic multi-head automata. Kutrib et. al. explored the computational power of one-way multi-head reversible automata [3] and reversible pushdown automata [4]. Pin et. al. applied algebraic theory to reversible automata to explore its computability [5]. We are interested in reversible automata because reversible automata are information preserving machines and information preserving machines help us to better analyze the way an automaton behaves. Moreover, the observation that loss of information results in heat dissipation further increases the interest in reversible automata. Kondacs et. al.[6] states that one-way reversible finite automata cannot accept all regular languages. Although Kutrib et. al.[3] have stated that one-way multi-head finite automaton with two heads can accept all uniletter regular languages, we still do not know whether one-way multi-head reversible finite automaton can accept all regular languages or not.

In this paper, our main objective is to find a one-way reversible finite automaton structure which accepts all regular language. In our quest for such a model, we were inspired by the Watson-Crick automata model introduced by Freund et. al.[7]. (Essentially Watson-Crick automata are finite automata having two independent heads working on double strands where the characters on the corresponding positions of the two strands are connected by a complementarity relation similar to the Watson-Crick complementarity relation. The movement of the heads although independent of each other is controlled by a single state). Details of several variants of non-deterministic Watson-Crick automata (AWK) have been explored in [8]. Deterministic Watson-Crick automata and their variants have been explicitly handled in [9]. Equivalence of subclasses of two-way Watson-Crick automata is discussed in [10]. A survey of Watson-Crick automata can be found in [11]. Research work regarding state complexity of Watson-Crick automata is reported in [12] and [13].

Sempere [14] worked on reversibility of Watson-Crick automata. He defined different kinds of regular reversibility in the model. Mainly, Sempere explored regular reversibility in the upper (lower) strand and in the double strand. Sempere did not introduce any formal model for reversible Watson-Crick automata. Thus, we introduce a new structure namely reversible Watson-Crick automaton (See Section 3), which is a reversible automaton that exploits the complementary relation and double strands properties of Watson-Crick automaton. The main claims of this paper are as follows:

- We show that the newly introduced reversible Watson-Crick automata structure can accept all regular languages (See Section 4).
- There is a reversible Watson-Crick automaton that accepts the language L = $\{\%w_1*x_1\%w_2*x_2...\%w_n*x_n | n \geq 0, w_i \in \{a,b\}^*, x_i \in \{a,b\}^*, \exists i \exists j : w_i = w_j, x_i \neq x_j\}$ which is not accepted by any multi-head deterministic finite automaton.(See Section 4)

## 2. BASIC TERMINOLOGY

The symbol V denotes a finite alphabet. The set of all finite words over V is denoted by $V^*$, which includes the empty word $\lambda$. The symbol $V^+=V^*- \{\lambda\}$ denotes the set of all non-empty words over the alphabet V. For $w \in V^*$, the length of w is denoted by $|w|$. Let $u \in V^*$ and $v \in V^*$ be two words and if there is some word $x \in V^*$, such that v=ux, then u is a prefix of v, denoted by $u \leq v$. Two words, u and v are prefix comparable denoted by $u \sim_p v$, if u is a prefix of v or vice versa.

### 2.1 Watson-Crick automata

A Watson-Crick automaton is a 6-tuple of the form $M=(V,\rho,Q,q_0,F,\delta)$ where V is an alphabet set, the symbol Q denotes the set of states, the complementarity relation $\rho \subseteq V \times V$ is similar to Watson-Crick complementarity relation, $q_0$ is the initial state and F⊆Q is the set of final states. The function δ contains a finite number of transition rules of the form $q\binom{w_1}{w_2} \rightarrow q'$, which denotes that the machine in state q parses $w_1$ in upper strand and $w_2$ in lower strand and goes to state q' where $w_1, w_2 \in V^*$. The symbol $\begin{bmatrix} w_1 \\ w_2 \end{bmatrix}$ is different from $\binom{w_1}{w_2}$. While $\binom{w_1}{w_2}$ is just a pair of strings written in that form instead of $(w_1,w_2)$, the symbol $\begin{bmatrix} w_1 \\ w_2 \end{bmatrix}$ denotes that the two strands are of same length i.e. $|w_1|=|w_2|$ and the corresponding symbols in two strands are complementarity in the sense given by the relation ρ. The symbol $\begin{bmatrix} V \\ V \end{bmatrix}_\rho = \{ \begin{bmatrix} a \\ b \end{bmatrix} \mid a, b \in V, (a, b) \in \rho \}$ and $WK_\rho(V) = \begin{bmatrix} V \\ V \end{bmatrix}_\rho^*$ denotes the Watson-Crick domain associated with V and ρ.

A transition in a Watson-Crick finite automaton can be defined as follows:

For $\binom{x_1}{x_2}, \binom{u_1}{u_2}, \binom{w_1}{w_2} \in \binom{V^*}{V^*}$ such that $\begin{bmatrix} x_1 u_1 w_1 \\ x_2 u_2 w_2 \end{bmatrix} \in WK_\rho(V)$ and q, q' ∈Q, $\binom{x_1}{x_2} q\binom{u_1}{u_2}\binom{w_1}{w_2} \Rightarrow \binom{x_1}{x_2}\binom{u_1}{u_2} q'\binom{w_1}{w_2}$ iff there is transition rule $q\binom{u_1}{u_2} \rightarrow q'$ in δ and $\overset{*}{\Rightarrow}$ denotes the transitive and reflexive closure of ⇒. The language accepted by a Watson-Crick automaton M is $L(M) = \{w_1 \in V^* | q_0 \begin{bmatrix} w_1 \\ w_2 \end{bmatrix} \overset{*}{\Rightarrow} q\begin{bmatrix} \lambda \\ \lambda \end{bmatrix}$, with q ∈ F, $w_2 \in V^*, \begin{bmatrix} w_1 \\ w_2 \end{bmatrix} \in WK_\rho(V)\}$.

## 2.2 Deterministic Watson-Crick automata

Czeizler et. al.[9] introduced the notion of determinism in Watson-Crick automata. Different notions of determinism of Watson-Crick automata are as follows:
1) weakly deterministic Watson-Crick automata(WDWK): Watson-Crick automaton is weakly deterministic if in every configuration that can occur in some computation of the automaton, there is a unique possibility to continue the computation, i.e. at every step of the automaton there is at most one-way to carry on the computation.
2) deterministic Watson-Crick automata(DWK): deterministic Watson-Crick automaton is Watson-Crick automaton for which if there are two transitions of the form $q\binom{u}{v} \rightarrow q'$ and $q\binom{u'}{v'} \rightarrow q''$ then $u \not\sim_p u'$ or $v \not\sim_p v'$.
3) strongly deterministic Watson-Crick automata(SDWK): strongly deterministic Watson-Crick automaton is a deterministic Watson-Crick automaton where the Watson-Crick complementarity relation is injective.

## 2.3 One-way reversible multi-head finite automata

The following definition of one-way reversible multi-head finite automata and deterministic multi-head finite automata are obtained from Kutrib et. al.[3]. In order to define one-way reversible multi-head finite automata Kutrib also defined one-way deterministic multi-head finite automata. We will also do so in order to define the reversible multi-head automata in the same manner as Kutrib.

**Definition 1:** A one-way deterministic k head finite automaton (1DFA(k)) is a system M= (Q, V,k,#, $, $q_0$, F,δ) where

1) Q is the finite set of states.

2) V is the finite set of input symbols.

3) k≥1 is the number of heads.

4) #∉V is the left end marker.

5) $∉V is the right end marker.

6) $q_0$∈Q is the initial state.

7) F⊆Q is the set of accepting states and

8) δ:Q×(V∪{#,$})$^k$→Q×{0,1}$^k$ is the partial transition function where 1 means to move the head one position to the right and 0 means to keep the head in the current position. Whenever δ(q',($a_1,a_2,…,a_k$))=(q,($d_1,d_2,…,d_k$)) is defined then $d_i$=0 if $a_i$=$, 1≤i≤k.

The configuration of a 1DFA(k) M= (Q, V,k, #, $, $q_0$, F, δ) at some time t≥0 is a triple $c_t$=(w,q,P) where w∈V$^*$, q∈Q is the current state and P=($p_1, p_2,…,p_k$)∈{0,1,…,|w|+1}$^k$ gives the current head position. If $p_i$=0, then the i$^{th}$ head is scanning the symbol #. If it satisfies 1≤$p_i$≤|w|, then the i$^{th}$ head is scanning the $p_i$th symbol of w in the upper strand and if $p_i$=|w|+1 then the i$^{th}$ head is scanning the end marker symbol $ at the end of w. The initial configuration for input is set to (w, $q_0$, (0,..,0)) where string in the upper strand is of the form #w$. During its course of computation M goes through a sequence of configurations. One step from a configuration to its successor configuration is denoted by ⊢. Let w=$x_1 x_2,…,x_n$, $x_0$=# and $x_{n+1}$=$, write (w,q,($p_1,p_2,...,p_k$))⊢(w,q',($p_1+d_1,p_2+d_2,…,p_k+d_k$)) if and only if δ(q,($a_{p_1}, a_{p_2},…, a_{p_k}$))=(q',($d_1,d_2,…,d_k$)) exists. The reflexive and transitive closure of ⊢ is denoted by ⊢$^*$. The language accepted by 1DFA(k) is precisely the set of words w such that there is some computation beginning with #w$ and the 1DFA(k) halts in an accepting state. A 1DFA(k) halts when the transition function is not defined for the current situation.

$L(M)=\{w \in V^* | (w,q_0,(0,\ldots,0)) \vdash^* (w,q,(p_1,p_2,\ldots,p_k)), q \in F,$ and M halts in $(w,q,(p_1,p_2,\ldots,p_k))\}$.

Based on Definition 1 of 1DFA(k) we define one-way multi-head reversible finite automaton in the following manner:

**Definition 2:** Let M be a 1DFA(k) and C be the set of all reachable configurations that occur in any computation of M beginning with an initial configuration and $(w,q,(p_1,p_2,\ldots,p_k)) \in C$ with $w=x_1x_2,\ldots,x_n$, $x_0=\#$ and $x_{n+1}=\$$.

M is said to be reversible, if the following two conditions are fulfilled:

1) For any two transitions $\delta(q',(a_1,a_2,\ldots,a_k))=(q,(d_1,d_2,\ldots d_k))$ and $\delta(q'',(a_1',a_2',\ldots,a_k'))=(q,(d_1',d_2',\ldots,d_k'))$ it holds that $(d_1,d_2,\ldots d_k)=(d_1',d_2',\ldots,d_k')$.

2) There is at most one transition of the form

$\delta(q',(x_{p_1-d_1}, x_{p_2-d_2},\ldots, x_{p_k-d_k}))=(q,(d_1,d_2,\ldots,d_k))$.

### 3. REVERSIBLE WATSON-CRICK AUTOMATA

In this Section, we introduce our model of reversible Watson-Crick automaton which is inspired by the structure of one-way reversible multi-head finite automaton defined by Kutrib et. al.[3](see Section 2) and two-way reversible multi-head finite automaton by Morita[2]. We replace the read only input tape of reversible multi-head finite automaton by a DNA double strand or a double tape. The content of the second tape is determined in the same manner as in Watson-Crick automaton using the complementarity relation $\rho:V \to V$ where $\rho$ is defined for every $x \in V$ and $\rho$ can be non-injective. We further introduce two independent reading heads one for each strand/tape. A formal definition of reversible Watson-Crick automaton is as follows:

**Definition 3:** A one-way reversible Watson-Crick automaton (1RWKA) is a system M= $(Q, V, \#, \$, q_0, F, \rho, \delta)$ where
1) Q is the finite set of states.
2) V is the finite set of input symbols.
3) $\# \notin V$ is the left end marker.
4) $\$ \notin V$ is the right end marker.
5) $q_0 \in Q$ is the initial state.
6) $F \subseteq Q$ is the set of accepting states
7) $\rho$ is the complementarity relation same as that defined in Watson-Crick automata i.e. $\rho:V \to V$ where $\rho$ is defined for every $x \in V$ and $\rho$ can be non-injective and
8) $\delta:Q \times (V \cup \{\#,\$\})^2 \to Q \times \{0,1\}^2$ is the partial transition function where 1 means to move the head one position to the right and 0 means to keep the head in the current position. Whenever $\delta(q',(a_1,a_2))=(q,(d_1,d_2))$ is defined then $d_i=0$ if $a_i=\$$ $1 \leq i \leq 2$.

To make the automaton reversible the following two restrictions are imposed on the partial transition function $\delta$;

1) for any two transitions $\delta(q',(a_1,a_2))=(q,(d_1,d_2))$ and $\delta(q'',(a_1',a_2'))=(q,(d_1',d_2'))$ it holds that $d_1'=d_1$ and $d_2'=d_2$.

2) for any two transitions $\delta(q',(a_1,a_2))=(q,(d_1,d_2))$ and $\delta(q'',(a_1',a_2'))=(q,(d_1,d_2))$ it holds that either $a_1 \neq a_1'$ or $a_2 \neq a_2'$ or $a_1 \neq a_1'$ and $a_2 \neq a_2'$.

Condition (1) ensures transitions resulting in the same state have to move the heads in the same way. Whereas condition (2) (termed by Morita[2] as the reversibility condition) ensures forward as well as backward determinism in the automaton. In this definition (Definition 3) we have replaced Kutrib et. al.'s reachability condition for reversible automaton by Morita's reversibility condition.

The configuration of a 1RWKA M= $(Q, V, \#, \$, q_0, F, \rho, \delta)$ at some time $t \geq 0$ is a quadruple $c_t=(w_1,w_2,q,P)$ where $w_1 \in V^*$ and $w_2 \in V^*$ where $w_2$ (the string in the second strand) is determined in the following manner:

If $w_1=\lambda$ then $w_2=\lambda$, if $w_1=x_1x_2,\ldots,x_n$ where $|w_1|=n$ and $x_i \in V$, $1 \leq i \leq n$ then $w_2=\rho(x_1)\rho(x_2),\ldots,\rho(x_n)$. The symbol $q \in Q$ is the current state and $P=(p_1, p_2) \in \{0,1,\ldots,|w_1|+1\}^2$ gives the current head position. If $p_1=0$, then the head on the upper strand is scanning the symbol # on the upper strand. If it satisfies $1 \leq p_1 \leq |w_1|$, then the upper head is scanning the $p_1$th symbol of $w_1$ in the upper strand and if $p_1=|w_1|+1$ then the upper head is scanning the end marker symbol $ at the end of $w_1$. The interpretation of the values of $p_2$ is similar to that of $p_1$. The only difference being $p_2$ denotes the position of the lower head placed on the lower strand and the lower head reads $w_2$. The initial configuration for input is set to $(w_1,w_2,q_0,(0,0))$ where string in the upper strand is of the form $\#w_1\$$ and the string in the lower strand is of the form $\#w_2\$$. During its course of computation M goes through a sequence of configurations. One step from a configuration to its successor configuration is denoted by $\vdash$. Let $w_1=c_1c_2,\ldots,c_n$, $c_0=\#$ and $c_{n+1}=\$$, $w_2$ is obtained from $w_1$ in a manner as described above and let it be of the form $w_2=b_1b_2,\ldots,b_n$, $b_0=\#$ and $b_{n+1}=\$$. We write $(w_1,w_2,q,(p_1,p_2)) \vdash (w_1,w_2,q',(p_1+d_1,p_2+d_2))$ if and only if $\delta(q,(c_{p_1}, c_{p_2}))=(q',(d_1,d_2))$ exists. The reflexive and transitive closure of $\vdash$ is denoted by $\vdash^*$.

The language accepted by 1RWKA is precisely the set of words $w_1$ such that there is some computation beginning with $\#w_1\$$ in the upper strand and $\#w_2\$$ in the lower strand. The string $w_2$ is obtained in the same manner as described in the beginning of this Section and 1RWKA halts in an accepting state. A 1RWKA halts when the transition function is not defined for the current situation.

$L(M)=\{w_1 \in V^*|(w_1,w_2,q_0,(0,0)) \vdash^* (w_1,w_2,q,(p_1,p_2)), w_2=\lambda$ if $w_1=\lambda$ otherwise $w_2= \rho(x_1)\rho(x_2),...,\rho(x_n)$ if $w_1=x_1x_2,...,x_n$ where $|w_1|=n$ and $x_i \in V$, $1 \leq i \leq n$, $q \in F$, and M halts in $(w_1,w_2,q,(p_1,p_2))\}$.

An input in the upper strand is accepted if and only if M halts in an accepting state; in all other cases it is rejected. That is if the computation halts in a rejecting state or if the computation runs into an infinite loop. In the case of infinite loop eventually all heads are stationary since the machine described is one-way.

A reversible Watson-Crick automaton is called a ***strongly reversible Watson-Crick automaton*** if the complementarity relation is injective in that particular automaton.

## 4. COMPUTATIONAL COMPLEXITY OF REVERSIBLE WATSON-CRICK AUTOMATA

In this Section, we discuss the computational power of reversible Watson-Crick automaton. We show that for every deterministic finite automaton which accepts a language L we can construct a reversible Watson-Crick automaton which accepts the same language L. In order to explain the construction, we first employ the rules of construction on a particular deterministic finite automaton which accepts the language $(a+b)^*a$ (from Kondacs et. al. work [6], we know L cannot be recognised by a reversible finite automaton with one head) and obtain the corresponding reversible Watson-Crick automaton which accepts the same language (See Example 1). Then in Theorem 1, we state the general proof for any deterministic finite automaton.

**Example 1:** Consider a deterministic finite automaton $M= (Q, V, q_0, F, \delta)$ which accepts the language $(a+b)^*a$, where $Q=\{q_0, q_1\}$, $V=\{a, b\}$, $F=\{q_1\}$ The transition function $\delta$ is as follows:

$\delta(q_0, a)=q_1$, $\delta(q_0, b)=q_0$, $\delta(q_1, a)=q_1$, $\delta(q_1, b)=q_0$.

The corresponding reversible Watson-Crick automaton that accepts $L=(a+b)^*a$ is obtained as follows:

$M'= (Q', V',\#, \$, q_0', F', \rho, \delta')$

$Q'=Q \cup \{q_0', q_f\}$, $F'=\{q_f\}$.

For every $x \in V$ we list the transitions associated with x and arrange them in any particular order and assign a number to each transition according to its position in the list. (See Table 1).

Table 1: The list of transitions

| x=a | x=b |
|---|---|
| $\delta(q_0, a)=q_1......a_1$ | $\delta(q_0, b)=q_0......b_1$ |
| $\delta(q_1, a)=q_1......a_2$ | $\delta(q_1, b)=q_0......b_2$ |

Now for each $x \in V$, if its list has n transitions, we introduce the symbols $x,x_1,...,x_n$ in V' and the elements $(x,x_1),(x,x_2),...,(x,x_n)$ in $\rho$.
Therefore in our particular case:
$V'=\{a, a_1, a_2, b, b_1, b_2\}$ and $\rho=\{(a, a_1), (a, a_2), (b, b_1), (b, b_2)\}$.
Now for each list of transitions the following steps are performed.
1) For a transition $\delta(q,x)=q'$ numbered $x_i$, we introduce the transition $\delta'(q,(x, x_i))=(q',(1,1))$ in $\delta'$. We repeat this step for every transition in the list.
So, for our particular example the following transitions are introduced in $\delta'$:
$\delta'(q_0,(a, a_1))=(q_1,(1, 1))$ and $\delta'(q_1,(a, a_2))=(q_1,(1, 1))$ for x=a and
$\delta'(q_0,(b, b_1))=(q_0,(1, 1))$ and $\delta'(q_1,(b, b_2))=(q_0,(1, 1))$ for x=b
In addition to the above transitions, the following transitions are also introduced in $\delta'$.
$\delta'(q_0',(\#, \#))=(q_0,(1,1))$: This transition ensures M' enters the start state $q_0$ of M and both upper and lower head of M' are on the next character after #.
$\delta'(q,(\$, \$))=(q_f,(0,0))$ for all $q \in F$: These transitions ensure that if M' in simulating M reaches the final state of M after consuming the string in both the strands, such that both its heads are on $ then M' goes to its accepting state and halts. Thus the transitions introduced in $\delta'$ for our particular example are $\delta'(q_0',(\#, \#))=(q_0,(1,1))$ and $\delta'(q_1,(\$, \$))=(q_f,(0,0))$.

Thus, all the transitions introduced in $\delta'$ according to the above stated rules are as follows:
$\delta'(q_0',(\#, \#))=(q_0,(1,1))$
$\delta'(q_0,(a, a_1))=(q_1,(1, 1))$

$\delta'(q_1,(a, a_2))=(q_1,(1, 1))$
$\delta'(q_0,(b, b_1))=(q_0,(1, 1))$
$\delta'(q_1,(b, b_2))=(q_0,(1, 1))$.
$\delta'(q_1,(\$, \$))=(q_f,(0,0))$

The set of states of M' i.e. $Q'=Q\cup\{q_0', q_f\}$. Thus in our Example $Q'=\{q_0', q_0, q_1, q_f\}$.
The set of final states of M' i.e. $F'=\{q_f\}$.

Suppose M accepts w. The complementarity relation $\rho$ of M' is so designed that the complementarity string w' of w guesses the transitions that M takes to accept w. Each position of w' guesses the transition that M takes on reading that particular position in w. Based on the sequence in w', M' simulates the transition sequence of M. As M accepts w, there is a sequence of transitions that takes M to its final state after consuming w. Thus, one of the many complementarity strings of w will rightly guess that particular sequence of transitions that enables M to accept w and for that particular complementarity string as M' simulates M based on the complementarity string of w; M' will reach the final state of M and both its heads will be on $. The transitions $\delta'(q,(\$,\$))=q_f$, $q\in F$, takes M' to its final state after consuming its input. Thus M' accepts w.

For better clarity of the above argument, let us consider an example of a w which M accepts;
Consider w=aba. As w ends in 'a' therefore M accepts w and $w\in L=(a+b)^*a$.
The transition that M applies to accept w are in the following sequence:
$\delta(q_0, a)=q_1$......$a_1$
$\delta(q_1, b)=q_0$......$b_2$
$\delta(q_1, a)=q_1$......$a_2$
i.e. the sequence of transitions applied by M to accept w in terms of the number assigned to each transition is $a_1b_2a_2$.
The possible complementarirty strings of w are as follows:
1) $a_1b_1a_1$
2) $a_2b_1a_1$
3) $a_1b_2a_1$
4) $a_2b_2a_1$
5) $a_1b_1a_2$
6) $a_2b_1a_2$
7) $a_1b_2a_2$
8) $a_2b_2a_2$.

So the 7$^{th}$ string matches with the sequence of transitions that M used to accept w. Thus there exist a complementarity string w"= $a_1b_2a_2$ which correctly guesses the transition sequence M employed to accept w. Thus M' on this complementarity string correctly simulates M to accept w.

The transitions employed by M' to accept w when guessed complementarity string is w" are as follows:
$\delta'(q_0',(\#, \#))=(q_0,(1,1))$
$\delta'(q_0,(a, a_1))=(q_1,(1, 1))$
$\delta'(q_1,(b, b_2))=(q_0,(1, 1))$.
$\delta'(q_1,(a, a_2))=(q_1,(1, 1))$
$\delta'(q_1,(\$, \$))=(q_f,(0,0))$.

For a string w, which M does not accept, there is no sequence of transitions that takes M to its final state after consumption of w. Thus, no matter what guess the complementarity strings of w make, M' while simulating M based on the complementarity strings of w will never reach the situation where both its head is on $ and M' is in a final state of M. Thus the transitions of the form $\delta'(q,(\$,\$))=q_f$, $q\in F$ cannot be applied to M'. As a result M' never reaches its final state $q_f$. Thus M' rejects w.

**Theorem 1:** For every deterministic finite automaton which accepts a language L, we can find a reversible Watson-Crick automaton which accepts the same language (L).

**Proof:** The proof of the above Theorem is in two parts. In the first part given a deterministic finite automaton M which accepts a language L, we give a construction to obtain a reversible Watson-Crick automaton M' from M and in the second part we show that M' accepts the same language as M.

**First Part**: Given a deterministic finite automaton M= $(Q, V, q_0, F,\delta)$. We construct a reversible Watson-Crick automaton M'= $(Q', V',\#, \$, q_0', F', \rho, \delta')$ from M in the following manner:

For every x∈V, we do the following steps:

1) We form a list of all the transitions in M involving x, where these transitions involving x are arranged in any particular order and each transition is assigned a number of the form $x_i$ based on its position in the list. i.e. a transition is assigned a number $x_i$, if the transition is the $i^{th}$ transition in the list for x∈V.

2) Let us suppose there are n transitions in the list, then introduce the symbols $x, x_1, ..., x_n$ in V' and the relations $(x, x_1), (x, x_2), ..., (x, x_n)$ in ρ.

3) For a transition δ(q,x)=q' having number $x_i$ associated with it, we introduce the transition $δ'(q, (x, x_i)) = (q', (1,1))$ in δ'.

Step 3 is repeated for every transition in the list.

Moreover the following transitions are also added to δ'.
1) $δ'(q_0', (\#, \#)) = (q_0, (1,1))$
2) $δ'(q, (\$, \$)) = (q_f, (0,0))$ for all q∈F
The set of states of M' i.e. Q'=Q∪{$q_0', q_f$}.
The set of final states of M' i.e. F'={$q_f$}.
The start state of M' is $q_0'$.

**Second Part:** In this part we show that M' constructed from M accepts the same language as M. The argument of this part is same as that of Example 1. We repeat the argument here so as to enable the reader to understand the proof of Theorem 1 without referring to Example 1.

Suppose M accepts w. The complementarity relation ρ of M' is so designed that the complementarity string w' of w guesses the transitions that M takes to accept w. Each position of w' guesses the transition that M takes on reading that particular position in w. Based on the sequence in w', M' simulates the transition sequence of M. As M accepts w, there is a sequence of transitions that takes M to its final state after consuming w. Thus, one of the many complementarity strings of w will rightly guess that particular sequence of transitions that enables M to accept w and for that particular complementarity string as M' simulates M based on the complementarity string of w; M' will reach the final state of M and both its heads will be on $. The transitions $δ'(q, (\$,\$))=q_f$, q∈F, takes M' to its final state after consuming its input. Thus M' accepts w.

For a string w, which M does not accept, there is no sequence of transitions that takes M to its final state after consumption of w. Thus, no matter what guesses the complementarity strings of w makes, M' while simulating M based on the complementarity string of w will never reach the situation where both its heads are on $ and M' is in a final state of M. Thus the transitions of the form $δ'(q, (\$,\$))=q_f$, q∈F cannot be applied to M'. As a result M' never reaches its final state $q_f$. Thus M will reject w.

**Corollary 1:** Reversible Watson-Crick automata can accept all regular languages.

**Proof:** From Theorem 1, we know that for every deterministic finite automaton there is a reversible Watson-Crick automaton which accepts the same language. For every regular language there is a deterministic finite automaton which accepts that language, thus for every regular languages there is a reversible Watson-Crick automaton that accepts it.

**Theorem 2:** The language L = {$\%w_1*x_1\%w_2*x_2...\%w_n*x_n | n \geq 0, w_i \in \{a,b\}^*, x_i \in \{a,b\}^*, \exists i \exists j : w_i = w_j, x_i \neq x_j$} is accepted by a reversible Watson-Crick automaton with non-injective complementarity relation.

**Proof:** Let, M= (Q, V, #, $, $q_0$, F, ρ, δ) be a reversible Watson-Crick automaton,
where V={a, b, $v_{m1}, v_{m2}$, %, *}, ρ={(a,a), (%,%), (%,$v_{m1}$), (%,$v_{m2}$), (b,b), (*,*)}, Q = {$q_0, q_1, q_2, q_3, q_4$}, F={$q_3$}, and we have the following transitions:

$δ(q_0,(\#,\#))=(q_0,(1,1))$, $δ(q_0,(\%,\%))=(q_0,(1,1))$, $δ(q_0,(a,a))=(q_0,(1,1))$, $δ(q_0,(b,b))=(q_0,(1,1))$, $δ(q_0,(*,*))=(q_0,(1,1))$, $δ(q_0,(\%,v_{m1}))=(q_1,(0,1))$, $δ(q_1,(\%,a))=(q_1,(0,1))$, $δ(q_1,(\%,b))=(q_1,(0,1))$, $δ(q_1,(\%,*))=(q_1,(0,1))$, $δ(q_1,(\%,\%))=(q_1,(0,1))$, $δ(q_1,(\%,v_{m2}))=(q_2,(1,1))$, $δ(q_2,(a,a))=(q_2,(1,1))$, $δ(q_2,(b,b))=(q_2,(1,1))$, $δ(q_2,(*,*))=(q_3,(1,1))$, $δ(q_3,(a,a))=(q_3,(1,1))$, $δ(q_3,(b,b))=(q_3,(1,1))$, $δ(q_3,(\%,\%))=(q_4,(0,0))$, $δ(q_3,(\%,\$))=(q_4,(0,0))$.

The above stated automaton works in the following manner:

The elements (%, $v_{m1}$) and (%, $v_{m2}$) of the complementarity relation ρ are used to guess the two substrings of the input string which has its w parts equal and x parts unequal. On finding these guessed substrings the automaton goes to state $q_2$. In state $q_2$, the automaton M checks to see whether the guessed substrings have their w parts equal or not. If the substrings do not have their w parts equal then the automaton halts in a non final state $q_2$ and rejects the input string. If the two guessed substring have their w parts equal then the automaton goes to state $q_3$. In state $q_3$, the automaton M checks whether the guessed substrings having their w parts equal have their x parts equal or not. If the x parts are equal then the automaton goes to state $q_4$. There are no transitions defined for state $q_4$ and it's a non final state. Thus, the automaton rejects the input string. If the x parts are unequal then the automaton halts in state $q_3$ which is a final state; thus the input string is accepted as the guessed substrings have their w parts equal and x parts not equal.

Consider a string s in L. One of the many complementarity strings of s will correctly guess the two substrings which have their w parts equal and x parts unequal and s will be accepted by the automaton M.

Now, consider a string s not in L. As s is not in L there are no two substrings of s whose w parts are equal and x parts are unequal. Therefore, no matter the guess made by any complementarity string of s for the location of two substrings of s they will never have their w parts equal and x parts unequal. So M rejects s. Thus, from the above stated arguments we conclude that M accepts L.

**Lemma 1:** The language $L = \{\%w_1*x_1\%w_2*x_2...\%w_n*x_n | n \geq 0, w_i \in \{a,b\}^*, x_i \in \{a,b\}^*, \exists i \exists j : w_i = w_j, x_i \neq x_j\}$ is not accepted by any deterministic multi-head finite automaton.

The proof of Lemma 1 is in Yao et. al.[15]

**Theorem 3:** $L_{1RWKA} - L_{DFA(k)} \neq \emptyset$, where $L_{1RWKA}$ is the set of all languages accepted by reversible Watson-Crick automata and $L_{DFA(k)}$ is the set of all languages accepted by multi-head deterministic finite automata.

**Proof:** From Theorem 2, we know that there is a reversible Watson-Crick automaton that accepts the language $L = \{\%w_1*x_1\%w_2*x_2...\%w_n*x_n | n \geq 0, w_i \in \{a,b\}^*, x_i \in \{a,b\}^*, \exists i \exists j : w_i = w_j, x_i \neq x_j\}$ and from Lemma 1 we know that $L = \{\%w_1*x_1\%w_2*x_2...\%w_n*x_n | n \geq 0, w_i \in \{a,b\}^*, x_i \in \{a,b\}^*, \exists i \exists j : w_i = w_j, x_i \neq x_j\}$ is not accepted by any deterministic multi-head finite automaton which proves the above Theorem.

**Corollary 2:** $L_{1RWKA} - L_{SDWK} \neq \emptyset$, where $L_{1RWKA}$ is the set of all languages accepted by reversible Watson-Crick automata and $L_{SDWK}$ is the set of all languages accepted by strongly deterministic Watson-Crick automata.

**Proof:** Czeizler et. al.[9] states that the computational power of strongly deterministic Watson-Crick automata and deterministic finite automata with two heads are same and from Theorem 3 we see $L_{1RWKA} - L_{DFA(k)} \neq \emptyset$ hence the above stated Corollary holds.

**Lemma 2:** For every multi-head reversible finite automaton with two heads which accepts a language L there is a strongly reversible Watson-Crick automaton which accepts the same language L.

**Proof:** Consider a reversible multi-head automaton with two heads $M = (V, Q, 2, \#, \$, q_0, F, \delta)$ the corresponding strongly reversible Watson-Crick automaton is obtained in the following manner:

We take $\rho$ to be the identity relation. The other sets such as the set of input symbols, set of states, the set of final states, start state and the end makers are same for both M and M'. The transition of $M' = (Q, V, \#, \$, q_0, F, \rho, \delta')$ are obtained from M as follows:

For a transition $\delta(q, (a, b)) = (q', (d_1, d_2))$, $a, b \in V$, $q, q' \in Q$ in $\delta$, the transition $\delta'(q, (a, b)) = (q', (d_1, d_2))$ is introduced in $\delta'$.

As the complementarity relation $\rho$ is the identity relation therefore the content of the second strand of M' is same as that of the first strand. Now the automaton M' simulates automaton M in the following way: Each movement of the first head of M is simulated by the head on the upper strand of M' and each movement of the second head of M is simulated by the head on the lower strand of M'. As the transitions, set of states, initial state, set of final states are identical for M and M', M' behaves in the same manner as M and accepts the same language.

**Lemma 3:** For every strongly reversible Watson-Crick automaton which accepts a language L there is a reversible multi-head finite automaton which accepts the same language L.

**Proof:** Given a strongly reversible Watson-Crick automaton $M = (V, Q, q_0, F, \delta, \#, \$, \rho)$ we can obtain a reversible multi-head automaton with two heads in the following manner:

As the Watson-Crick automaton is strongly reversible therefore $\rho$ is an injective relation and so $\rho^{-1}(x)$, $x \in V$ can be uniquely determined.

The corresponding reversible multi-head automaton with two heads has its alphabet set, set of final states, set of states and the start state same as M. Thus $M' = (V, Q, F, \delta', \#, \$, q_0)$. The transitions of M' are obtained from M as follows:

For $\delta(q, (a, x)) = (q', (d_1, d_2))$, $a, x \in V$, $q, q' \in Q$ in $\delta$. We introduce $\delta'(q, (a, \rho^{-1}(x))) = (q', (d_1, d_2))$ in $\delta'$. As $\rho$ is injective there is only one such element and that element must be present in the upper strand at that particular position where we obtain x in the lower strand because contents of the lower strand is obtained by applying the relation ρ to each symbol of the word w in the upper strand. Thus the first head of the two head reversible multi-head automaton simulates the head of the upper strand of the reversible Watson-Crick automaton whereas the second head simulates the head on the lower strand. Thus M' behaves exactly as M. So M' accepts the same language as M.

**Theorem 4:** Strongly reversible Watson-Crick automata and multi-head reversible finite automata with two heads have the same computational power.

**Proof:** Lemma 2 tells us that for every multi-head reversible finite automaton with two heads which accepts a language L there is a strongly reversible Watson-Crick automaton which accepts the same language L. From Lemma 3 we know that for every strongly reversible Watson-Crick automaton which accepts a language L there is a reversible multi-head finite automaton which accepts the same language L. Thus we can say that strongly reversible Watson-Crick automata and multi-head reversible finite automata with two heads have the same computational power.

**Theorem 5**: The set of languages accepted by strongly reversible Watson-Crick automata is a proper subset of the set of languages accepted by reversible Watson-Crick automata.

**Proof:** Any strongly reversible Watson-Crick automaton is also reversible Watson-Crick automaton. Thus every language accepted by strongly reversible Watson-Crick automata are accepted by reversible Watson-Crick automata. Moreover from Theorem 4, we know that strongly reversible Watson-Crick automata and multi-head reversible finite automata with two heads have the same computational power and it has already been stated by Kutrib et.al.[3] that set of languages accepted by multi-head reversible finite automata is a proper subset of set of languages accepted by multi-head deterministic finite automata. Thus there is no strongly reversible Watson-Crick finite automaton which accept the language L = $\{\%w_1*x_1\%w_2*x_2...\%w_n*x_n | n \geq 0, w_i \in \{a,b\}^*, x_i \in \{a,b\}^*, \exists i \exists j : w_i = w_j, x_i \neq x_j\}$ but from Theorem 2, we see that a reversible Watson-Crick automaton can accept the language L, which proves the Theorem.

## 5. CONCLUSION

In this paper we have introduced a new model of reversible finite automaton and we show that the new model namely reversible Watson-Crick automata can accept all regular languages. We further showed that this new model accepts a language which is not accepted by any multi-head deterministic finite automata. Moreover, we compared the computational power of our reversible model with other existing one-way reversible and deterministic finite automata in literature.